\documentclass[10pt,english,comsoc,conference]{IEEEtran}
\usepackage[T1]{fontenc}
\usepackage[latin9]{inputenc}
\usepackage{geometry}
\usepackage{amsmath}
\usepackage{amsthm}
\usepackage{amssymb}
\usepackage{graphicx}

\makeatletter

\providecommand{\tabularnewline}{\\}

  \theoremstyle{definition}
  \newtheorem{defn}{\protect\definitionname}
  \theoremstyle{plain}
  \newtheorem{lem}{\protect\lemmaname}
  \theoremstyle{plain}
  \newtheorem{thm}{\protect\theoremname}

\usepackage[T1]{fontenc}
\usepackage[latin9]{inputenc}
\usepackage{array}
\usepackage{multirow}
\usepackage{amsmath}
\usepackage{amsthm}
\usepackage{amssymb}
\usepackage{graphicx}
\usepackage{geometry}
\makeatother

\usepackage{babel}
\providecommand{\definitionname}{Definition}
\providecommand{\lemmaname}{Lemma}
\providecommand{\theoremname}{Theorem}

\begin{document}
\title{
A Mean-Field Stackelberg Game Approach for Obfuscation Adoption in Empirical Risk Minimization 
}



\author{

\IEEEauthorblockN{Jeffrey Pawlick } 
  \IEEEauthorblockA{New York University Tandon School of Engineering \\Department of Electrical and Computer Engineering \\ Email: jpawlick@nyu.edu} 

\and 

\IEEEauthorblockN{Quanyan Zhu} 
  \IEEEauthorblockA{New York University Tandon School of Engineering \\Department of Electrical and Computer Engineering \\ Email: quanyan.zhu@nyu.edu} 
\thanks{This work is partially supported by the grant CNS-1544782,  EFRI-1441140  and  SES-1541164  from National Science Foundation.}
}
\maketitle
\begin{abstract}
Data ecosystems are becoming larger and more complex due to online
tracking, wearable computing, and the Internet of Things. But privacy
concerns are threatening to erode the potential benefits of these
systems. Recently, users have developed obfuscation techniques that
issue fake search engine queries, undermine location tracking algorithms,
or evade government surveillance. Interestingly, these techniques
raise two conflicts: one between each user and the machine learning
algorithms which track the users, and one between the users themselves.
In this paper, we use game theory to capture the first conflict with
a Stackelberg game and the second conflict with a mean field game.
We combine both into a dynamic and strategic bi-level framework which
quantifies accuracy using empirical risk minimization and privacy
using differential privacy. In equilibrium, we identify necessary
and sufficient conditions under which 1) each user is incentivized
to obfuscate if other users are obfuscating, 2) the tracking algorithm
can avoid this by promising a level of privacy protection, and 3)
this promise is incentive-compatible for the tracking algorithm.\end{abstract}


\section{Introduction }

We often hear that data is the new oil. On the Internet, websites
sell user information to third-party trackers such as advertising
agencies, social networking sites, and data analytic companies \cite{mayerthird-party2012}.
In the Internet of things (IoT), devices such as smartwatches include
accelerometers, heart rate sensors, and sleep trackers that measure
and upload data about users' physical and medical conditions \cite{swansensor2012}.
At a larger scale, smart grid and renewable energy also stand to benefit
from developments in networks of sensors and actuators \cite{baheti2011cyber}.

While these technologies promise positive impacts, they also threaten
privacy. Specifically, wearable computing and IoT devices collect
sensitive information such as health and location data \cite{internet2015}.
In addition, the pervasiveness of tracking allows learners to infer
habits and physical conditions over time. For instance, tracking algorithms
may predict ``a user's mood; stress levels; personality type; bipolar
disorder; demographics'' \cite{peppetregulating2014}. These are
unprecedented degrees of access to user information.

Interestingly, users have recently begun to take privacy into their
own hands using basic tools for \emph{obfuscation}. Obfuscation is
``the deliberate addition of ambiguous, confusing, or misleading
information to interfere with surveillance and data collection''
\cite{brunton2015obfuscation}. In a signal processing sense, obfuscation
provides noise.

\begin{figure}
\begin{centering}
\includegraphics[width=0.85\columnwidth]{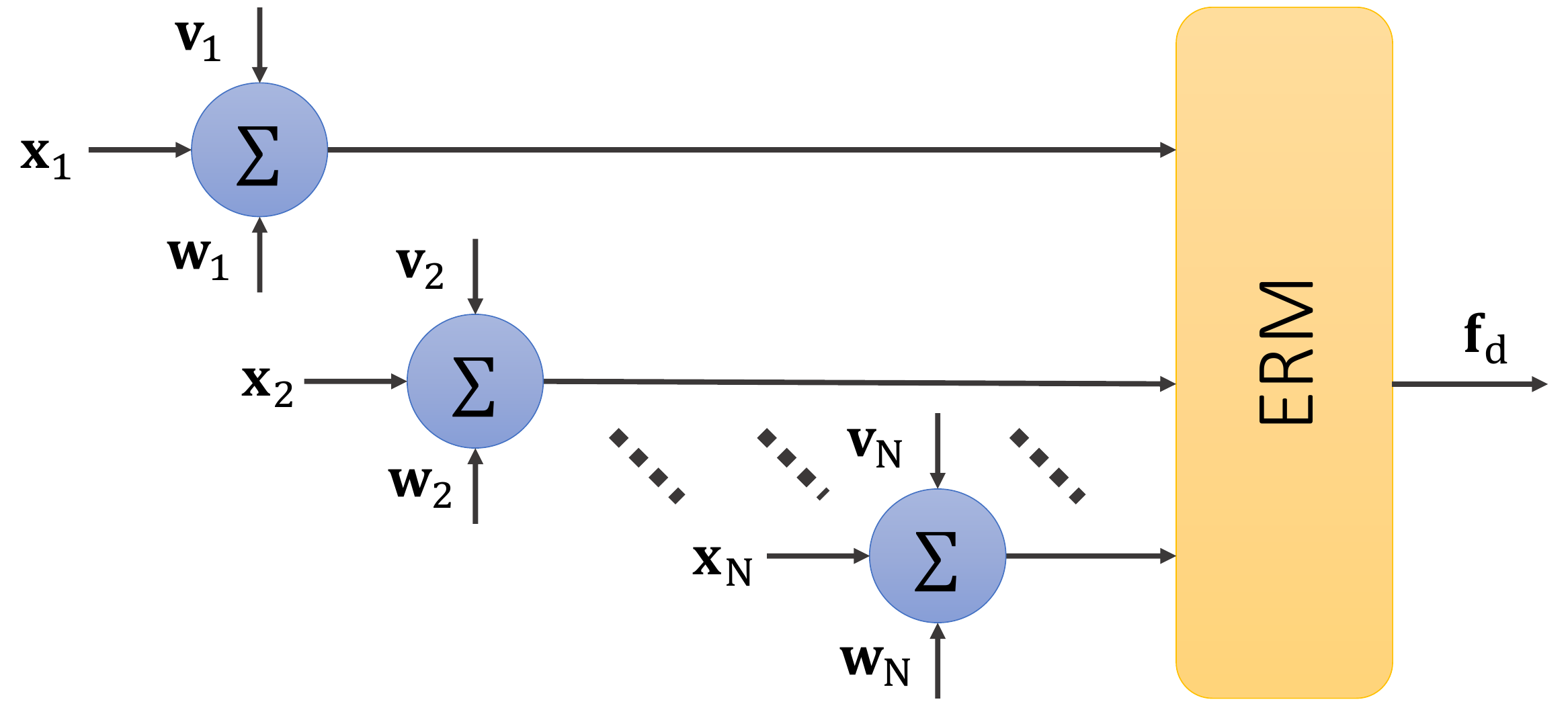} 
\par\end{centering}

\caption{\label{fig:dataFlow}Data flow in the obfuscation-tracking model.
Users $1,\ldots,N$ have data $\mathbf{x}_{i}$ with labels $y_{i}.$
They add noise $\mathbf{v}_{i}\sim\mathcal{V}_{i}$ to $\mathbf{x}_{i},$
and the learner promises noise $\mathbf{w}_{i}\protect\overset{\text{i.i.d.}}{\sim}\mathcal{W}.$
Noise degrades accuracy but improves privacy. The users and the learner
have misaligned incentives.}
\end{figure}

Finn and Nissenbaum describes two recent obfuscation technologies:
\emph{CacheCloak} and \emph{TrackMeNot} \cite{brunton2015obfuscation}.
\emph{TrackMeNot} is a browser extension that generates randomized
search queries in order to prevent trackers from assembling accurate
profiles of their users \cite{Howe2009}. In the realm of the IoT,
\emph{CacheCloak} provides a way for a user to access location-based
services without revealing his or her exact geographical position
\cite{meyerowitzhiding2009}. The app predicts multiple possibilities
for the path of a user, and then retrieves location-based information
for each path. An adversary tracking the requests is left with many
possible paths rather than a unique one. 
of these obfuscation technologies see them as a way for users to resist
tracking and put pressure on machine learning algorithms to guarantee
some privacy protection.

\emph{In this paper, we use game theory to identify conditions under
which the threat of user obfuscation motivates machine learners to
promise privacy protection.} We construct a bi-level framework to
model this interaction. In the user level, a large number of users
play a mean-field game (MFG) (\emph{c.f.} \cite{chan2015bertrand})
to decide whether to use obfuscation. In the learner level, a machine
learner plays a Stackelberg game (SG) \cite{vonstackelbergmarktform1934}
to decide whether to promise some level of privacy protection in order
to avoid obfuscation by the users.

Related work includes research in \emph{privacy markets}, in which
a learner pays users to report data truthfully \cite{ghosh2015selling,xiao2013privacy}.
Our paper differs by allowing the learner to promise privacy protection.
In \cite{chessa2015short,chessa2015game}, users play a multiple person,
prior-commitment game, which determines how much they obfuscate. In
these papers, the learner calculates the average of a dataset, while
in our framework a learner can use empirical risk minimization to
compute more general statistics. Finally, \cite{pawlick2016stackelberg}
considers a Stackelberg game, but unlike the present paper, it does
not include a mean-field interaction among the users. This interaction
captures a cascading effect by which many users may rapidly adopt
obfuscation technology.

\section{Model\label{sec:ERM_DP_Mod} }

Figure \ref{fig:dataFlow} depicts an interaction between a set of
users $i\in\mathbb{S}=\left\{ 1,\ldots,N\right\} $ and a learner
$L.$ Users submit possibly-perturbed data to $L,$ and $L$ releases
a statistic or predictor $\mathbf{f}_{d}$ of the data. Assume that
the data generating process is a random variable $\mathcal{Z}$ with
a fixed but unknown distribution. Denote the realized data by $\mathbf{z}_{i}\overset{\text{i.i.d.}}{\sim}\mathcal{Z},\,i\in\mathbb{S}.$
Each data point is composed of a feature vector $\mathbf{x}_{i}\in\mathbb{R}^{d}$
and a label $y_{i}\in\left\{ -1,1\right\} .$ The goal of the learner
$L$ is to predict $y_{i}$ given $\mathbf{x}_{i},$ based on the
trained classifier or predictor $\mathbf{f}_{d}.$

We investigate whether it is advantageous for $L$ to promise some
level of privacy protection in order to avoid user obfuscation \footnote{$L$ can accomplish this by collecting data at low resolution. This
is consistent with the spirit of DP, in which a learner publishes
$\epsilon_{p}.$}. $L$ adds noise with the same variance to each data point $\mathbf{x}_{i}.$
For $i\in\mathbb{S},$ $k\in1,\ldots,d,$ $L$ draws $w_{i}^{\left(k\right)}\overset{\text{i.i.d}}{\sim}\mathcal{W},$
where $\mathcal{W}$ is a mean-zero Gaussian random variable with
standard deviation $\sigma_{L}.$ While DP often considers Laplace
noise, we use Gaussian noise for reasons of mathematical convenience.
Knowing $\sigma_{L},$ each user adds noise $v_{i}^{\left(k\right)}\overset{\text{i.i.d.}}{\sim}\mathcal{V}_{i},$
$k\in1,\ldots,d,$ where $\mathcal{V}_{i}$ is Gaussian with variance
$\sigma_{S}^{i}.$ It is also convenient to define $\bar{\sigma}_{S}^{2}=\frac{1}{N}\sum_{i=1}^{N}(\sigma_{S}^{i})^{2},$
the average variance of the perturbations of every user, and $(\bar{\sigma}_{S}^{-i})^{2}=\frac{1}{N}\sum_{j=1}^{N}(\sigma_{S}^{j})^{2}-\frac{1}{N}(\sigma_{S}^{i})^{2},$
the average variance of the perturbations of every\emph{ }user \emph{other
than} $i.$ The perturbed data points are given by $\mathbf{\tilde{x}}_{i}=\mathbf{x}_{i}+\mathbf{v}_{i}+\mathbf{w}_{i},$
$i\in\mathbb{S}.$

\subsection{Empirical Risk Minimization\label{sub:Empirical-Risk-Minimization}}

Empirical risk minimization (ERM) refers to one popular family of
machine learning. In ERM, $L$ calculates a value of an output $\mathbf{f}_{d}\in\mathbf{F}$
that minimizes the empirical risk, \emph{i.e.}, the total penalty
due to imperfect classification of the realized data. Define a \emph{loss
function} $l\left(\mathbf{\tilde{z}}_{i},\mathbf{f}\right),$ which
expresses the penalty due to a single perturbed data point $\mathbf{\tilde{z}}_{i}$
for the output $\mathbf{f}.$ $L$ obtains $\mathbf{f}_{d}$ given
by Eq. \ref{eq:thetaD}, where $\rho\geq0$ is a constant and $R\left(\mathbf{f}\right)$
is a regularization term to prevent overfitting: 
\begin{equation}
\mathbf{f}_{d}=\underset{\mathbf{f}\in\mathbf{F}}{\arg\min}\,\rho R\left(\mathbf{f}\right)+\frac{1}{N}\underset{i=1}{\sum^{N}}l\left(\mathbf{\tilde{z}}_{i},\mathbf{f}\right),\label{eq:thetaD}
\end{equation}

Expected loss provides a measure of the accuracy of the output of
ERM. Let $\mathbf{f^{*}}$ denote the $\mathbf{f}$ which minimizes
the expected loss for unperturbed data: 
\begin{equation}
\mathbf{f^{*}}=\underset{\mathbf{f}\in\mathbf{F}}{\text{argmin}}\,\mathbb{E}\left\{ \rho R\left(\mathbf{f}\right)+l\left(\mathcal{Z},\mathbf{f}\right)\right\} .
\end{equation}
In Definition \ref{def:ERMacc}, $\mathbf{f^{*}}$ forms a reference
to which the expected loss of the perturbed classifier $\mathbf{f}_{d}$
can be compared. 
\begin{defn}
\label{def:ERMacc}($\epsilon_{g}$-Accuracy) Let $\mathbf{f}_{d}$
and $\mathbf{f^{*}}$ denote the perturbed classifier and the classifier
which minimizes expected loss, respectively. Let $\epsilon_{g}$ be
a positive scalar. We say that $\mathbf{f}_{d}$ is $\epsilon_{g}$-accurate
if it satisfies 
\begin{equation}
\mathbb{E}\left\{ \rho R\left(\mathbf{f}_{d}\right)+l\left(\mathcal{Z},\mathbf{f}_{d}\right)\right\} \leq\mathbb{E}\left\{ \rho R\left(\mathbf{f^{*}}\right)+l\left(\mathcal{Z},\mathbf{f^{*}}\right)\right\} +\epsilon_{g}.\label{eq:utilDif}
\end{equation}

\end{defn}
Lemma \ref{lem:accConst} obtains $\epsilon_{g}$ as a function of
the obfuscation levels. 
\begin{lem}
\label{lem:accConst}(Accuracy Level) If $L$ perturbs with variance
$\sigma_{L}^{2},$ user $i\in\mathbb{S}$ perturbs with $(\sigma_{S}^{i})^{2},$
and the other users perturb with $(\bar{\sigma}_{S}^{-i})^{2},$ then
the difference $\epsilon_{g}$ in expected loss between the perturbed
classifier and the population-optimal classifier is on the order of
\[
\epsilon_{g}\left(\sigma_{L},\bar{\sigma}_{S}^{-i},\sigma_{S}^{i}\right)\propto\frac{1}{\rho^{2}N}\left(\sigma_{L}^{2}+\frac{N-1}{N}\left(\bar{\sigma}_{S}^{-i}\right)^{2}+\frac{1}{N}\left(\sigma_{S}^{i}\right)^{2}\right).
\]

\end{lem}

\subsection{Differential Privacy\label{sub:DP}}

Using differential privacy (DP), a machine learning agent promises
a bound $\epsilon_{p}$ on the maximum information leaked about an
individual. Let $\mathcal{A}\left(*\right)$ denote an algorithm and
$D$ denote a database. Let $D'$ denote a database that differs from
$D$ by only one entry (\emph{e.g.}, the entry of the user under consideration).
Let $c$ be some set among all possible sets $C$ in which the output
of the algorithm $\mathcal{A}$ may fall. Then Definition \ref{def:DP}
quantifies privacy using the framework of DP \cite{chaudhuri2011differentially,dwork2006differential}. 
\begin{defn}
\label{def:DP}($\epsilon_{p}$-Privacy) - An algorithm $\mathcal{A}\left(B\right)$
taking values in a set $C$ provides $\left(\epsilon_{p},\delta\right)$-differential
privacy if, for all $D,$ $D'$ that differ in at most one entry,
and for all $c\in C,$ 
\begin{equation}
\mathbb{P}\left\{ \mathcal{A}\left(D\right)\in c\right\} \leq\exp\left\{ \epsilon_{p}\right\} \mathbb{P}\left\{ \mathcal{A}\left(D'\right)\in c\right\} +\delta.\label{eq:DP}
\end{equation}

\end{defn}
For a cryptographically-small $\delta,$ the degree of randomness
determines the privacy level $\epsilon_{p}$. Lower values of $\epsilon_{p}$
correspond to more privacy. That randomness is attained through the
noise added in the forms of $\mathcal{V}$ and $\mathcal{W}.$ 
\begin{lem}
(Privacy Level) If $L$ adds noise with variance $\sigma_{L}^{2}$
and user $i\in\mathbb{S}$ perturbs with variance $(\sigma_{S}^{i})^{2},$
then the user obtains differential privacy level $\epsilon_{p}\in(0,1)$
on the order of 
\begin{equation}
\epsilon_{p}\left(\sigma_{L},\sigma_{S}^{i}\right)\propto\left(\sigma_{L}^{2}+\left(\sigma_{S}^{i}\right)^{2}\right)^{-1/2}.
\end{equation}

\end{lem}

\subsection{Bi-Level Game\label{sub:Bi-Level-Game}}

\begin{figure}
\begin{centering}
\includegraphics[width=0.72\columnwidth]{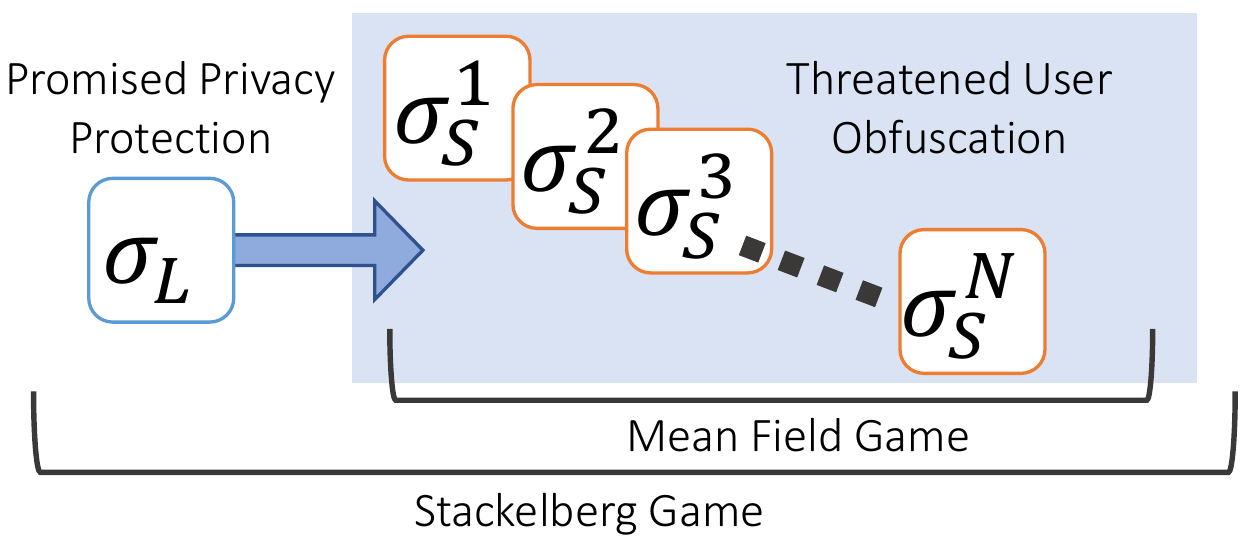} 
\par\end{centering}

\caption{\label{fig:biLevel}Bi-level structure of the strategic interaction.
Users may adopt obfuscation technologies in a cascading manner. This
is modeled by an MFG. To avoid this, a learner can proactively add
noise to their data. His interaction with the users in modeled by
an SG. }
\end{figure}
Let $\mathbb{R}_{M}$ denote a subset $[0,M]$ of the non-negative
real numbers, and let $M$ be arbitrarily large\footnote{This rigorously deals with large perturbation variances. }.
Let $\sigma_{L}\in\mathbb{R}_{M}$ denote the noise variance added
by the learner. If the users are not satisfied with this level of
privacy protection, they may add noise with variances $\sigma_{S}^{i}\in\mathbb{R}_{M},$
$i\in\mathbb{S}.$

Define a utility function by $U_{L}:\,\mathbb{R}_{M}^{2}\to\mathbb{R}$
such that $U_{L}(\sigma_{L},\bar{\sigma}_{S})$ gives the utility
that $L$ receives for using noise $\sigma_{L}^{2}$ while the users
add an average noise of $\bar{\sigma}_{S}^{2}.$ Also define utility
functions $U_{S}^{i}:\,\mathbb{R}_{M}^{3}\to\mathbb{R}$ such that
user $i\in\mathbb{S}$ receives utility $U_{S}^{i}(\sigma_{L},\bar{\sigma}_{S}^{-i},\sigma_{S}^{i})$
for obfuscating with variance $(\sigma_{S}^{i})^{2}$ while the other
users obfuscate with average variance $(\bar{\sigma}_{S}^{-i})^{2}$
and $L$ perturbs with $\sigma_{L}^{2}.$ $U_{L}$ and $U_{S}^{i},$
$i\in\mathbb{S},$ are given by

\[
U_{L}\left(\sigma_{L},\bar{\sigma}_{S}\right)=A_{L}\exp\left\{ -\epsilon_{g}\left(\sigma_{L},\bar{\sigma}_{S}^{-i},\sigma_{S}^{i}\right)\right\} -C_{L}\mathbf{1}_{\left\{ \sigma_{L}>0\right\} },
\]
\begin{multline*}
U_{S}^{i}\left(\sigma_{L},\bar{\sigma}_{S}^{-i},\sigma_{S}^{i}\right)=A_{S}^{i}\exp\left\{ -\epsilon_{g}\left(\sigma_{L},\bar{\sigma}_{S}^{-i},\sigma_{S}^{i}\right)\right\} \\
-P_{S}^{i}\left(1-\exp\left\{ -\epsilon_{p}\left(\sigma_{L},\sigma_{S}^{i}\right)\right\} \right)-C_{S}^{i}\mathbf{1}_{\left\{ \sigma_{S}^{i}>0\right\} },
\end{multline*}
where $A_{L}$ (resp. $A_{S}^{i}$) gives the maximum benefit to the
learner (resp. to each user) for output accuracy, $P_{S}^{i}$ gives
the maximum privacy loss to each user, and $C_{L}$ (resp. $C_{S}^{i}$)
gives the flat cost of perturbation for the learner (resp. to each
user).

\subsection{Equilibrium Requirements\label{sub:Equilibrium-Requirements}}

Chronologically, $L$ first promises perturbation $\sigma_{L},$ and
then the users choose obfuscation $\sigma_{S}^{i},$ $i\in\mathbb{S}.$
The solution, however, proceeds backwards in time.

\subsubsection{Mean-Field Game}

Given the promised $\sigma_{L},$ the group of users plays a MFG in
which each user best responds to the average perturbation of the other
users\footnote{This is a strategic interaction, because each user would prefer to
protect her own privacy while making use of accurate data from the
other users.}. Consider symmetric utility functions for the users\footnote{That is, $A_{S}^{i}=A_{S},$ $P_{S}^{i}=P_{S},$ and $C_{S}^{i}=C_{S},$
$i\in\mathbb{S}.$}. Let $BR_{S}:\,\mathbb{R}_{M}\to\mathbb{R}_{M}$ denote a best response
function, such that 
\begin{equation}
BR_{S}\left(\bar{\sigma}_{S}^{-i}\,|\,\sigma_{L}\right)=\underset{\sigma_{S}^{i}\in\mathbb{R}_{M}}{\arg\max}\,U_{S}^{i}\left(\sigma_{L},\bar{\sigma}_{S}^{-i},\sigma_{S}^{i}\right)
\end{equation}
gives the set of best responses for user $i\in\mathbb{S}$ to the
average perturbation $\bar{\sigma}_{S}^{-i}$ of the other users,
given that the learner has promised $\sigma_{L}^{2}.$ Then the equilibrium
of the MFG is $\sigma_{S}^{1*}=\sigma_{S}^{2*}=\ldots=\sigma_{S}^{N*}$
(which is also equal to $\bar{\sigma}_{S}^{*}$) which satisfies the
fixed-point equation 
\begin{equation}
\bar{\sigma}_{S}^{*}\in BR_{S}\left(\bar{\sigma}_{S}^{*}\,|\,\sigma_{L}\right).\label{eq:fp}
\end{equation}
Now define a mapping $\Gamma:\,\mathbb{R}_{M}\to\mathbb{R}_{M}$ such
that $\Gamma(\sigma_{L})$ gives the $\bar{\sigma}_{S}^{*}$ which
satisfies Eq. (\ref{eq:fp}) given\footnote{We will apply a selection criteria to ensure there is only one $\bar{\sigma}_{S}^{*}.$ }
$\sigma_{L}.$ We say that, by promising $\sigma_{L},$ $L$ \emph{induces}
$\bar{\sigma}_{S}^{*}=\Gamma(\sigma_{L}).$

\subsubsection{Stackelberg Game}

Since $L$ promises $\sigma_{L}^{2}$ before the users obfuscate,
$L$ is a Stackelberg leader, and the users are collectively a Stackelberg
follower which plays $\Gamma(\sigma_{L})$. The optimality equation
for $L$ is 
\begin{equation}
\sigma_{L}^{*}\in\underset{\sigma_{L}\in\mathbb{R}_{M}}{\arg\max}\,U_{L}\left(\sigma_{L},\Gamma\left(\sigma_{L}\right)\right).\label{eq:stackOpt}
\end{equation}

\begin{defn}
\label{def:pbne}(Perfect Bayesian Nash Equilibrium) A perfect Bayesian
Nash equilibrium (PBNE) (\emph{c.f.}, \cite{fudenberg1991game}) of
the overall game is $(\sigma_{L}^{\dagger},\sigma_{S}^{1\dagger},\sigma_{S}^{2\dagger},\dots,\sigma_{S}^{N\dagger})$
such that $\bar{\sigma}_{S}^{\dagger}=\sigma_{S}^{1\dagger}=\sigma_{S}^{2\dagger}=\dots=\sigma_{S}^{N\dagger},$
and 
\begin{equation}
\bar{\sigma}_{S}^{\dagger}=\Gamma\left(\sigma_{L}^{\dagger}\right)=BR_{S}\left(\bar{\sigma}_{S}^{\dagger}\,|\,\sigma_{L}^{\dagger}\right),\label{eq:pbneMFG}
\end{equation}
\begin{equation}
\sigma_{L}^{\dagger}\in\underset{\sigma_{L}\in\mathbb{R}_{M}}{\arg\max}\,U_{L}\left(\sigma_{L},\Gamma\left(\sigma_{L}\right)\right).\label{eq:pbneSG}
\end{equation}

\end{defn}

\section{Mean Field Game Analysis\label{sec:Mean-Field-Game}}

\begin{figure}
\begin{centering}
\includegraphics[width=0.96\columnwidth]{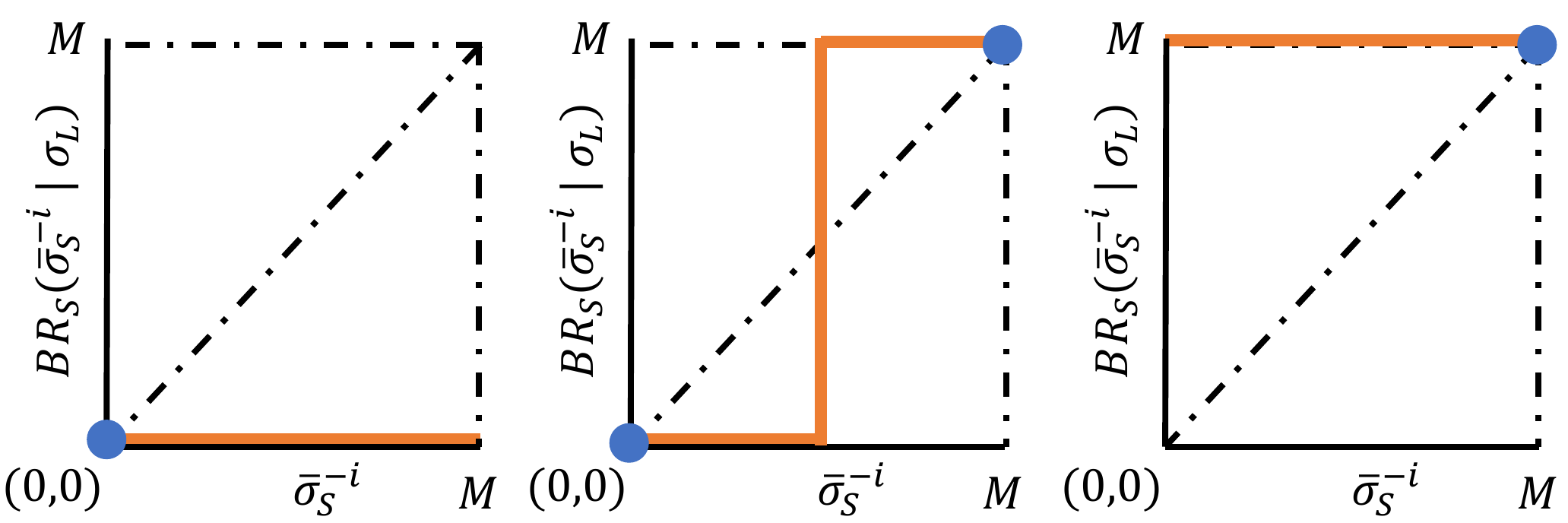} 
\par\end{centering}

\caption{\label{fig:br}Best response mappings (orange) for each user $i$
against the other users $-i.$ MFG equilibria occur at the intersections
(blue circles) of the mappings with the identity mapping.}
\end{figure}

First, Lemma \ref{lem:BR} solves for $BR_{S}.$ 
\begin{lem}
\label{lem:BR}(Best Response) Define $\mathbf{AC}(\sigma_{L},\bar{\sigma}_{S}^{-i})\triangleq A_{S}\exp\{-\epsilon_{g}(\sigma_{L},\bar{\sigma}_{S}^{-i},0)\}+C_{S}$
and $\mathbf{P}(\sigma_{L})\triangleq P_{S}(1-\exp\{-\epsilon_{p}(\sigma_{L},0)\}).$
Then $BR_{S}$ is given by 
\[
BR_{S}\left(\bar{\sigma}_{S}^{-i}\,|\,\sigma_{L}\right)=\begin{cases}
0, & \text{if }\mathbf{P}\left(\sigma_{L}\right)<\mathbf{AC}\left(\sigma_{L},\bar{\sigma}_{S}^{-i}\right)\\
M, & \text{if }\mathbf{P}\left(\sigma_{L}\right)>\mathbf{AC}\left(\sigma_{L},\bar{\sigma}_{S}^{-i}\right)\\{}
[0,M], & \text{if }\mathbf{P}\left(\sigma_{L}\right)=\mathbf{AC}\left(\sigma_{L},\bar{\sigma}_{S}^{-i}\right)
\end{cases}.
\]

\end{lem}
Figure \ref{fig:br} depicts Lemma \ref{lem:BR}. Users with low privacy
sensitivity (left) never obfuscate, while users with high privacy
sensitivity (right) always obfuscate. Importantly, users with moderate
privacy sensitivity (center) cascade: each user $i$ obfuscates if
$\bar{\sigma}_{S}^{-i}$ is high. 
Theorem \ref{thm:MFGeq} states that the MFG equilibria occur at the
fixed points of the best response mappings\footnote{Mixed strategies are omitted due to limited space.}. 
\begin{thm}
\label{thm:MFGeq}(MFG Equilibrium) Given a promised privacy protection
level $\sigma_{L}^{\dagger},$ Eq. (\ref{eq:pbneMFG}) is satisfied
by the symmetric strategies $\sigma_{S}^{\dagger1}=\ldots=\sigma_{S}^{\dagger N}=\bar{\sigma}_{S}^{\dagger},$
where $\bar{\sigma}_{S}^{\dagger}=\Gamma\left(\sigma_{L}^{\dagger}\right)$

\[
=\begin{cases}
0, & \text{if }\mathbf{P}\left(\sigma_{L}\right)<\mathbf{AC}\left(\sigma_{L},M\right)<\mathbf{AC}\left(\sigma_{L},0\right)\\
\{0,M\} & \text{if }\mathbf{AC}\left(\sigma_{L},M\right)\leq\mathbf{P}\left(\sigma_{L}\right)\leq\mathbf{AC}\left(\sigma_{L},0\right)\\
M & \text{if }\mathbf{AC}\left(\sigma_{L},M\right)<\mathbf{AC}\left(\sigma_{L},0\right)<\mathbf{P}\left(\sigma_{L}\right)
\end{cases}.
\]

\end{thm}
In the middle case, $U_{S}^{i}$ is higher for $\bar{\sigma}_{S}^{\dagger}=0$
than for $\bar{\sigma}_{S}^{\dagger}=M.$ Therefore, we select $\bar{\sigma}_{S}^{\dagger}=0$
and write $\Gamma(\sigma_{L})=M\mathbf{1}_{\{\mathbf{P}(\sigma_{L})>\mathbf{AC}(\sigma_{L},0)\}}.$

\section{Stackelberg Game\label{sec:Stackelberg-Game}}

Next, $L$ chooses $\sigma_{L}$ in order to maximize $U_{L}(\sigma_{L},\Gamma(\sigma_{L})).$

\subsection{Status Quo Equilibrium}

Lemma \ref{lem:trivialSG} gives a solution in which $L$ does not
perturb. 
\begin{lem}
\label{lem:trivialSG}(Status Quo SG Solution) If $P_{S}-C_{S}<A_{S},$
then $\Gamma_{L}(0)=0.$ In this case, the optimal $\sigma_{L}^{\dagger}=0,$
for which $L$ receives his maximum possible utility: $U_{L}(0,0)=A_{L}.$ 
\end{lem}
$P_{S}-C_{S}<A_{S}$ holds if users are willing to suffer a total
loss of privacy in order to obtain complete accuracy. We have called
this the \emph{status quo} because it seems to represent the current
preferences of many users.

\subsection{Equilibrium Outside of the Status Quo}

Consider $P_{S}-C_{S}>A_{S}.$ Define $\tau\in\mathbb{R}_{M}$ such
that $\mathbf{P}(\tau)=\mathbf{AC}(\tau,0).$ By promising to perturb
with at least $\tau,$ $L$ is able to induce $\Gamma(\tau)=0,$ \emph{i.e.},
to make it incentive-compatible for the users to not obfuscate. But
we must analyze whether promising $\tau$ is incentive-compatible
for $L.$ Since the analytical expression for $\tau$ is cumbersome,
define an approximation $\hat{\tau}>\tau,$ where $\hat{\tau}^{2}=1/\ln\{P_{S}/(P_{S}-C_{S})\}.$
Next, define $\kappa\triangleq1/(\rho^{2}N).$ Then $U_{L}\left(\sigma_{L},\Gamma\left(\sigma_{L}\right)\right)$
is 
\[
\approx\begin{cases}
0, & \text{if }\sigma_{L}^{2}=0\\
-C_{L}, & \text{if }0<\sigma_{L}^{2}<\hat{\tau}^{2}\\
A_{L}\exp\left\{ -\kappa\sigma_{L}^{2}\right\} -C_{L}, & \text{if }\hat{\tau}^{2}\leq\sigma_{L}^{2}<M
\end{cases}.
\]
$U_{L}$ is maximized by either $0$ or $\hat{\tau}$ according to
Theorem \ref{thm:SGeq}. 
\begin{thm}
\label{thm:SGeq}(SG Equilibrium) For $P_{S}-C_{S}>A_{S},$ the perturbation
promise which satisfies Eq. (\ref{eq:pbneSG}) is 
\begin{equation}
\sigma_{L}^{\dagger}=\begin{cases}
0, & \text{if }\frac{1}{\rho^{2}N}>\ln\left\{ \frac{A_{L}}{C_{L}}\right\} \ln\left\{ \frac{P_{S}}{P_{S}-C_{S}}\right\} \\
\hat{\tau}, & \text{if }\frac{1}{\rho^{2}N}<\ln\left\{ \frac{A_{L}}{C_{L}}\right\} \ln\left\{ \frac{P_{S}}{P_{S}-C_{S}}\right\} 
\end{cases}.
\end{equation}

\end{thm}
Theorem \ref{thm:SGeq} shows that high costs $C_{S}$ of user perturbation
incentivize $L$ to promise privacy protection, because users easily
decide not to obfuscate. On the other hand, high privacy sensitivity
$P_{S}$ decreases $L$'s incentive to add noise. Somewhat surprisingly,
high accuracy sensitivity $A_{L}$ leads $L$ to promise privacy protection\footnote{Accuracy sensitivity increases his sensitivity to user obfuscation.}. 

\begin{table}
\caption{\label{tab:results}Equilibrium Results of the Bi-Level Game}

\centering{}%
\begin{tabular}{|c|c|c|}
\hline 
Parameter Regime  & $\bar{\sigma}_{S}^{\dagger}$  & $\sigma_{L}^{\dagger}$\tabularnewline
\hline 
\hline 
1) $P_{S}-C_{S}<A_{S}$  & $0$  & $0$\tabularnewline
\hline 
2) $P_{S}-C_{S}>A_{S}$ $\bigcap$ $\frac{1}{\rho^{2}N}>\ln\left\{ \frac{A_{L}}{C_{L}}\right\} \ln\left\{ \frac{P_{S}}{P_{S}-C_{S}}\right\} $  & \multirow{1}{*}{$M$}  & \multirow{1}{*}{$0$}\tabularnewline
\hline 
3) $P_{S}-C_{S}>A_{S}$ $\bigcap$ $\frac{1}{\rho^{2}N}<\ln\left\{ \frac{A_{L}}{C_{L}}\right\} \ln\left\{ \frac{P_{S}}{P_{S}-C_{S}}\right\} $  & \multirow{1}{*}{$0$}  & \multirow{1}{*}{$\hat{\tau}$}\tabularnewline
\hline 
\end{tabular}
\end{table}

\subsection{Summary of Results\label{sub:Summary-of-Results}}

Table \ref{tab:results} summarizes the results of the overall game.
The equilibrium strategies $\bar{\sigma}_{S}^{\dagger}$ and $\sigma_{L}^{\dagger}$
satisfy Definition \ref{def:pbne}. Equilibrium 1 is the \emph{status
quo} equilibrium in which users submit unperturbed data and $L$ does
not protect it. This equilibrium achieves complete accuracy at the
cost of complete loss of privacy. In Equilibrium 2, users obfuscate
as much as possible. $L$ lacks incentive to promise privacy protection,
so he does not perturb. He receives zero utility, making machine learning
useless. Equilibrium 3 is the best equilibrium. In this scenario,
\emph{the threat} of user obfuscation convinces $L$ to promise privacy
protection $\sigma_{L}^{2}=\hat{\tau}^{2}.$ The users accept this
level, and do not adopt obfuscation.

\section{Discussion of Results}

Privacy skeptics argue that users are not willing to pay for privacy
protection. This is captured by $P_{S}-C_{S}<A_{S},$ which leads
to Equilibrium 1. But as obfuscation technologies such as \emph{TrackMeNot}
\cite{Howe2009} and \emph{CacheCloak} \cite{meyerowitzhiding2009}
continue to develop, the cost $C_{S}$ of obfuscation will decrease,
and the awareness $P_{S}$ of privacy concerns will increase. Both
will lead to $P_{S}-C_{S}>A_{S}.$ In Equilibrium 3, obfuscation motivates
the learner to promise some level of privacy protection. Nevertheless,
technologists should be careful about the effects of obfuscation.
In the case of Equilibrium 2, users perturb their data as much as
possible, but this only decreases the opportunities for meaningful
analysis and discourages machine learning. Our work lays a foundation
for identifying the scenarios in which adoption of obfuscation is
beneficial.

 \bibliographystyle{plain}
\bibliography{PDoSjBib}

\end{document}